\documentclass{v23windows}

\usepackage{xspace}       %

\newcommand{\Al}{$^{26}${Al}\xspace}

\newcommand{\Msol}{M\ensuremath{_\odot}\xspace}

\newcommand{\apj}{Astrophys. J. }%
\newcommand{\apjl}{Astrophys. J. Lett. }%
\newcommand{\apjs}{Astrophys. J. Suppl. Ser. }%
\newcommand{\apss}{Astroph.J.\&Sp.Sci. }%
\newcommand{\aap}{Astron. Astrophys. }%
\newcommand{\aj}{Astron. J. }%
\newcommand{\mnras}{Mon. Notices Royal Astron. Soc. }%
\newcommand{\nar}{New Astron. Rev. }%
\newcommand{\nat}{Nature }%
\newcommand{\physrep}{Phys.~Rep. }%
\newcommand{\Photo}{\includegraphics[height=35mm]{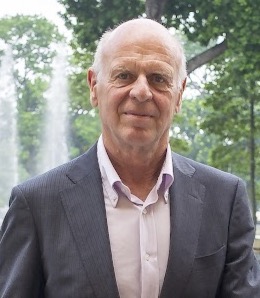}}

\begin{document}
\vspace*{4cm}
\title{Tracing the ejecta from cosmic nucleosynthesis}

\author{ Roland Diehl }

\address{Max Planck Institut f\"ur extraterrestrische Physik and TU M\"unchen, D-85748 Garching, Germany}

\maketitle\abstracts{
Long-lived radioactive by-products of nucleosynthesis provide an opportunity to trace the flow of ejecta away from its sources for times beyond where ejecta can be seen otherwise.
Gamma rays from such radioactive decay in interstellar space can be measured with space-borne telescopes. A prominent useful example is \Al with a radioactive decay time of one My.
Such observations have revealed that typical surroundings of massive stars are composed of large cavities, extending to kpc sizes.
Implications are that material recycling into new stars is twofold: rather direct as parental clouds are hosts to new star formation triggered by feedback, and
more indirect as these large cavities merge with ambient interstellar gas after some delay. 
Kinematic measurements of hot interstellar gas carrying such ejecta promises important measurements complementing stellar and dense gas kinematics.}

\section{Introduction}

Nucleosynthesis occurs in the universe within stars and during stellar explosions, with a small contribution outside stars from cosmic ray interactions in interstellar space. The original material for first such nucleosynthesis was set by the abundances resulting from big bang nucleosynthesis. Thereafter, nucleosynthesis happens in objects connected to and following from the formation of stars.
These sources of nucleosynthesis in the universe eject their products into their surroundings. Eventually, these ejecta mix with other cosmic gas. As new stars eventually are formed from such ambient cosmic gas, condensing and collecting under the force of gravity, such nucleosynthesis material finds its way into stars, where nuclear fusion reactions occur again, but now on enriched stellar material. This is the cycle of matter, as we call it, implementing successive enrichment of cosmic gas with products of nucleosynthesis. 
The successive enrichment of cosmic gas with products of nucleosynthesis is called \emph{cosmic chemical evolution}\cite{Diehl:2023}, as we mostly follow it through observing abundances of the chemical elements. No chemical reactions are involved, rather nuclear fusion reactions among atomic nuclei. Therefore, a more appropriate notion would  be \emph{cosmic evolution of the isotopic abundances}.

In this cycle of matter, the pathway of ejecta towards the next generation of forming stars is a significant step, and a challenge for astrophysics. Ejecta can be directly observed as they are heated or cooling down, within remnants or afterglows of the nucleosynthesis event. A prominent example is SN1987A, where we can witness ejecta dynamics over decades of expansion of the explosion remnant \cite{Cigan:2019,Jones:2023}. Such opportunities fade after several 10,000~years, when remnant material is too diluted or cold and faint to be discriminated from background. But what we learned in any case is that ejecta distribution cannot simply be modelled as isotropic diffusion or ballistic expansion. 
The reason is that (i) the complex interstellar medium is the agent receiving the ejecta, transporting it, and converting it to star-forming gas; and (ii) the interstellar environment differs for many of the different sources of nucleosynthesis.

The interstellar gas itself is now understood as highly dynamic and multi-phase medium, energized by massive stars in general, with their winds and supernovae, although rare events such as thermonuclear supernovae, and less-violent events such as planetary nebulae or novae also play a role. The energy injected by such explosions firsdt generates bubbles or cavities around the source \cite{Weaver:1977}, and eventually is converted into turbulence \cite{Krumholz:2005,Padoan:2016,Vazquez-Semadeni:2015}, which then propagates according to plasma interactions, i.e. mostly without collisions of atoms and thermal or ionization states that could reach an equilibrium. Therefore, interstellar plasma processes must be dealt with following the processes for each plasma component, from high-energy cosmic rays through electrons and ions down to cold atoms and molecules, phase transitions providing another challenge here. 

Common methods of studying the interstellar medium employ magnetic hydrodynamics equations in numerical treatments \cite{Breitschwerdt:2004,Vazquez-Semadeni:2015,Krumholz:2018}, where Euclidian and Lagrangian binned approaches are complemented by smooth-particle hydrodynamics. In the latter, virtual particles of adopted mass in the range of multiple solar masses are traced in the hydrodynamic treatment, assuming they could be representative for an ensemble of gas and its electrons and nuclei (or molecules).
The realism of such treatment of interstellar medium on all relevant scales of time and space is still to be established, in spite of much progress to reproduce some characteristics of the interstellar medium as we know it. 
Sub-dominant components such as ejecta from nucleosynthesis sources may, however, not be represented properly, as their injection itself occurs into non-representative extreme outskirts of interstellar medium, and propagation of matter may be more complex as propagation of injected energy happens in those same regions from the boundary conditions of violent injection.

Therefore, observations of ejecta flow from the sources into the greater ambient interstellar medium are of prominent value.
Radioactive isotopes provide a unique opportunity herein, as the radioactive decay imprints a decay law that modifies intensities of observable emission in a predictable and controlled way, independent of any plasma conditions, with fading intensity as ejecta age and thus merge with ambient interstellar medium. 
Radioactive \Al is such a tracer, with a radioactive decay time of 1.04~My, and characteristics gamma-ray line emission at an energy of 1808.63~keV \cite{Endt:1998}.
This emission has been detected first in our Galaxy in 1978 \cite{Mahoney:1982}, and has been measured in great detail by the Compton Gamma-Ray Observatory mission \cite{Gehrels:1993} of NASA (1991-2000) and by the INTEGRAL mission \cite{Winkler:2003} of ESA (2002 - today); latest results are shown in Figure~1. 

\begin{figure} 
\centering
  \includegraphics[width=0.8\textwidth]{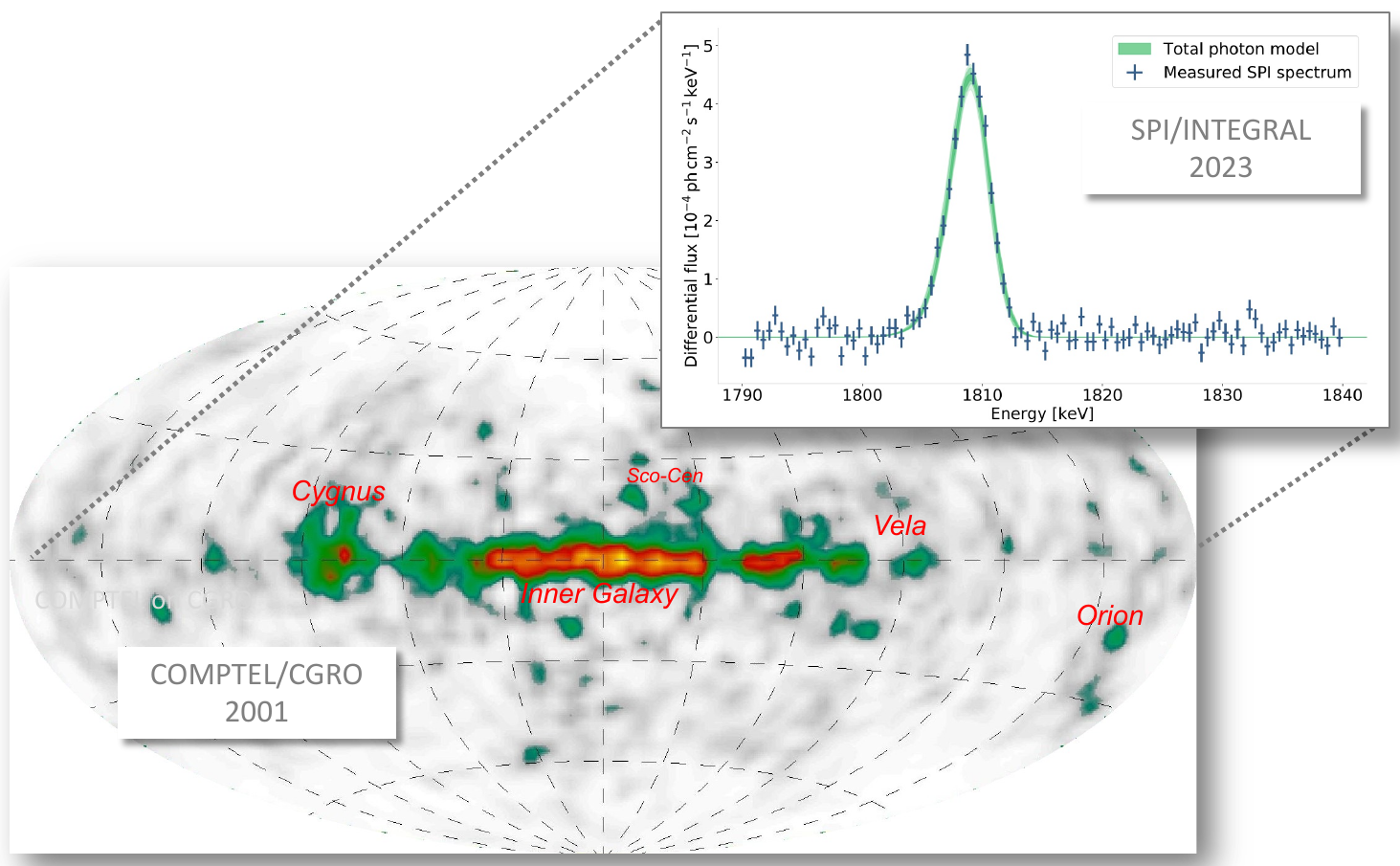}
  \caption{The \Al emission results from the COMPTEL and SPI measurements. The COMPTEL image shows extended and irregular emission along the plane of the Galaxy, with prominent emission from, e.g., the Cygnus region. The SPI spectrum shows a narrow but slightly-broadened line at the expected energy. (\it {Citations see text})}
  \label{fig:Al-map-spectrum}
\end{figure} 

The main CGRO-COMPTEL result was an all-sky image of \Al emission, obtained from 9 years of observations through the maximum-entropy deconvolution method \cite{Pluschke:2001c}. This image showed that the \Al emission extends along the disk of the Galaxy, with a rather bright inner region, and secondary emission features in the Cygnus region, and weak but interesting features in the Vela-Carina, Orion, and Scorpius-Centaurus regions. The main conclusion from this rather extended but irregular emission was that massive stars and their supernovae are the most plausible sources, as both AGB stars and in particular novae would contribute with much larger numbers of individual sources and thus result in a smoother spatial distribution than observed \cite{Prantzos:1996a}.
The main SPI/INTEGRAL result (beyond generally confirming the all-sky distribution of emission \cite{Bouchet:2015}) was extraction of kinematic information from detailed spectroscopy of the \Al line, measuring Doppler shifts of the line that most likely reflect the large-scale Galactic rotation \cite{Kretschmer:2013}.
These \Al observations and their astrophysical context have been addressed  in detail elsewhere \cite{Diehl:2018f,Diehl:2021}. 
Here we discuss recent findings and implications as relevant for ejecta propagation from nucleosynthesis sources.

\section{Results from \Al $\gamma$-ray spectroscopy}
\subsection{INTEGRAL/SPI line spectroscopy}\label{subsec:integral}

With INTEGRAL's spectrometer instrument SPI\cite{Vedrenne:2003,Roques:2003,Diehl:2018}, it has been possible to measure \Al emission with an energy resolution that enables kinematic analysis, as the line is resolved to a precision so that Doppler shifts and broadenings appear in line position and width. The achieve precision extends down to velocities below 100~km~S$^{-1}$.
After a first indication of the signature from large-scale galactic rotation \cite{Diehl:2006d}, the longitude-velocity signature was measured all along the inner Galaxy directions ($-60^\circ \leq l \leq 60^\circ$) \cite{Kretschmer:2013}. 
As the inferred \Al velocities were found to exceed Galactic-rotation velocities measured through CO (molecular gas) or masers (stars in specific local conditions) by about 200~km~s$^{-1}$, it seemed clear that fresh nucleosynthesis ejecta from mainly massive stars \cite{Prantzos:1996a} propagate differently than cold interstellar gas or stars.
The interpretation proposed from these results \cite{Krause:2014,Krause:2015} is that the \Al sources on average are surrounded by asymmetric large cavities, where ejecta travelling into the general direction of galactic rotation find a longer pathway in tenuous gas than ejecta travelling against galactic rotation. This was understood to arise from the genesis of cavities in inter-arm regions of the Galaxy, such that the density gradient away from the massive-star clusters at the leading edges of spiral arms favours a wider cavity extent away from spiral arms. 
This asymmetry with respect to the sources of ejecta thus imprints a biased velocity signature, as \Al decays during its propagation: much of the forward-propagating \Al decays while in the cavity, whereas much of the backward-propagating \Al already reached the cavity walls where it is decelerated to typical interstellar velocities before its decay. 
The longitude-velocity signature of \Al as compared to other Galactic tracers thus establishes on observational grounds  the existence of such asymmetric superbubbles with elongations up to kpc in our Galaxy.
Note that in other galaxies, where we have a face-on viewing aspect, such large and elongated superbubbles have been clearly seen in HI emission \cite{Bagetakos:2011} and in molecular gas \cite{Watkins:2023}. 

A specific such superbubble can also be found nearby. The Orion region\cite{Bally:2008} with its Orion-Eridanus bubble \cite{Heiles:1999} provides a favourable geometry for observations of how interstellar medium appears locally: The parental Orion molecular clouds are at a distance of about 450~pc, stellar groups including most-massive young stars can be well observed \cite{Chen:2020} and constrained in richness and kinematics, and the interstellar medium between the Sun and the Orion clusters features the well-observed large cavity called \emph{the Eridanus bubble}, which extends from the Orion clouds almost to the local bubble around the Sun \cite{Burrows:1993}. 
The geometry of this cavity is suggestive as one incarnation of the superbubbles inferred from large-scale Galactic \Al measurements as discussed above. 
It is encouraging that first analysis of the \Al emission from the Orion region \cite{Siegert:2016c} indicate a blue-shifted \Al line, which would be consistent with expectations of \Al ejecta propagating from the Orion star clusters more freely towards the Sun than in opposite direction and towards the Orion molecular clouds. 

\subsection{Interpreting the measurements: a bottom-up model}\label{subsec:integral}

Starting from the above interpretation of massive stars and their supernovae as dominant sources, one may exploit models of these objects for stars across their entire mass range to assemble the evolution of their outputs in energy and mass across time after their formation. 
Assuming coeval formation of an entire star cluster, such \emph{population synthesis} models have been built, to describe how the cluster surroundings are fed with ionizing starlight, with  kinetic energy from winds and supernovae, and with nucleosynthesis ejecta \cite{Voss:2009}. 
Herein, massive-star modeling from formation up to the supernova stage is used in tabulated time sequence data, and core-collapse supernova energetics and nucleosynthesis are used. The numbers of objects per stellar mass is derived from an initial-mass distribution through appropriate sampling, so that from the most-massive star\cite{Weidner:2006} to stars of 8~\Msol, the entire relevant mass range is populated and then tracked from stellar evolution, accumulating their respective outputs per time bin.

We recently updated \cite{Pleintinger:2020,Siegert:2023} (see references therein) our earlier model with latest insights from modeling massive-star evolution\cite{Limongi:2006a,Limongi:2018} including binaries\cite{Brinkman:2019}, and supernovae\cite{Sukhbold:2016,Janka:2016}, 
also including explodability\cite{Janka:2012,Pejcha:2020} criteria.
Then we enhanced this towards constructing a full Galactic bottom-up model for \Al emission as observable: 
We use a set of different large-scale models (spiral arm and exponential disk models, with disk scale height as additional parameter)  for the parental star-forming material in the Galaxy. We randomly sample hypothetical clusters of massive stars at locations herein and with a randomly-chosen age, until the total flux from \Al in the model equals the total Galactic flux measured with SPI. 
Herein we apply the above population synthesis model for each cluster, and also assign the \Al from each cluster a spatial distribution that corresponds to a superbubble size proportional to the age and total mass of the cluster. 
Ray tracing is then applied to the \Al amounts that are thus placed at different locations within the Galaxy, accounting for radioactive decay since ejection.
In this way, we obtain a synthetic bottom-up map of \Al emission, as constrained by the total measured flux in the \Al line and from the model inputs with their specific parameters.

A next step was to fit these hypothetical maps of \Al emission to raw spectra as measured with SPI for the sum of its observations through nearly 20 years of observations, thus using the SPI data without any further imaging deconvolution. In this way, one avoids the largely ill-determined deconvolution methods that turn our set of $\gamma$-ray spectra to a sky image, and rather uses the data space itself for a quantitative comparison of model and observed data. In order to account for the statistical variations of individual model incarnations, we sample typically 100 realisations for each plausible parameter choice, which allows to derive a statistical uncertainty of each such parameter within the model.  We can then perform a maximum-likelihood analysis to adjust the various components and parameters of the bottom-up model on our measured $\gamma$-ray data, and derive a fit quality which allows ranking of different such models.
For a reference, also the COMPTEL \cite{Pluschke:2001c} and SPI \cite{Bouchet:2015} spatial maps of \Al emission (which were constructed from imaging deconvolution methods without using any astronomical or astrophysical information)  have been fitted to the same data. We find that the bottom-up map approaches the fit quality of these observations-only cases (which naturally provide an even better fit) close enough to allow interpretation of the model parameter results as being based on a plausible and valid model.

The overall best-fitting model map is shown in Figure~2 (top). 
It has been constructed with a large-scale Galactic star formation distribution of a four-arm spiral, and a scale height of 700~pc.
The stellar-evolution model tat turned out best is the FRANEC model not including stellar rotation \cite{Limongi:2006a}.
As for supernova explosions, the explodability criteria\cite{Pejcha:2020} that avoid explosions with ejecta above 25~\Msol have been found best\cite{,Pleintinger:2020}. 
For these parameter choices, the absolute-flux constraint then resulted in an intensity scaling that would correspond to a star formation rate of 8~\Msol~y$^{-1}$. 
Finally, the intensity scaling has been adjusted to match those of the dynamic range shown in Figure~1, and a Gaussian smearing of 2$^\circ$ converts the resolution of the bottom-up map to the instrumental imaging resolution of SPI.
 
These model parameter values defining the best fit to observations are however not unique\cite{Pleintinger:2020,Siegert:2023}, as uncertainties and cross-correlated parameter variations occur. In particular, we observe fit residuals being larger in high-latitude regions and also towards the anticenter. 

We further enhance this bottom-up model by introducing knowledge about massive star groups in the solar system vicinity: Catalogues\cite{Melnik:1995,Melnik:2017} have been used are rather complete out to distances beyond 3--4~kpc. We found that this catalogue-based \Al amission accounts for 20-30\% of the total \Al emission alone, confirming the above result of the importance of nearby sources and the closest spiral-arm segments. Having realistic locations of the prominent nearby \Al sources then still improves the fit slightly, in particular residuals in the Cygnus and Scorpius-Centaurus regions are reduced,  beyond a model randomly sampling from the large-scale Galaxy model\cite{Pleintinger:2020}. Nevertheless, fit residuals at high latitudes and in the anticenter region remain (see discussion below).

\begin{figure} 
\centering
  \includegraphics[width=0.6\textwidth]{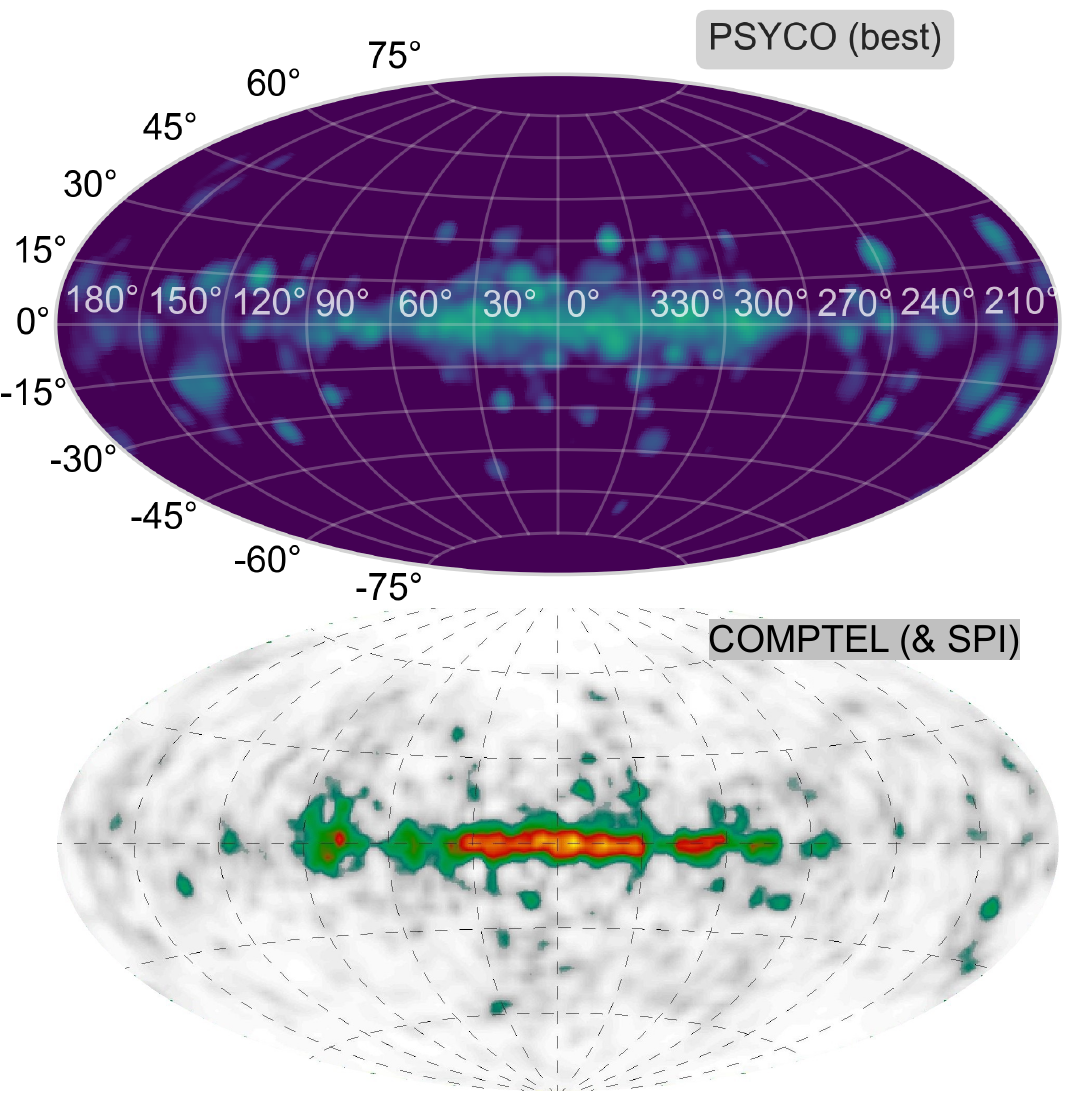}
  \caption{The\Al emission bottom-up model from population synthesis (above) versus  the COMPTEL image (below).  For details see text.}
  \label{fig:Al-map-scomparison}
\end{figure} 

\section{Discussion and Conclusions}

By establishing a bottom-up model for \Al from massive-star groups in our Galaxy, which uses stellar evolution and supernova model knowledge and population synthesis of such groups, we find that we can represent rather well the appearance of \Al images as derived from observations. 
We see that the \Al measurements can be understood and seen to be consistent with a model that assigns the source of \Al to groups of massive stars as they are created and evolve throughout our Galaxy. 
The agreement is satisfactory, within the uncertainties of the measurement (statistics, instrumental and analysis systematics) and the model (limited knowledge about the massive star groups in our Galaxy; systematic limitations and parameter uncertainties within the nucleosynthesis modeling of massive stars; systematic uncertainties of how \Al may be distributed within the subsequently-formed cavities around the massive star groups). 
Fitting a parametrized variety of such models to observed data, and analysing the model variants and their parameters, we face limitations from these systematics uncertainties and from degeneracies of parameter impacts in the spatial morphology of the \Al emission. 

Nevertheless,  the observed data support importance of spiral-arm structure from in particular the nearest spiral arms, and rather disfavour models with enhanced star formation in the central regions (out to 2-3~kpc) of the Galaxy. The scale height favoured by our model fits turns out to be significantly larger (about 700~pc)  than the scale height of observed massive stars (about 150~pc). Remaining residuals in particular at high galactic latitudes indicate that our superbubble models may be particularly unrealistic for the important nearby massive-star groups that contribute a large part of the \Al signal in data. This also may indicate an important role for superbubble blowouts into the Galactic halo, as also emerged from hydrodynamic simulations of a representative galaxy\cite{Fujimoto:2019,Rodgers-Lee:2019}. 

Beyond these rather consistently-found morphological characteristics of Galactic \Al emission, some inconsistency in absolute normalization appears, as indicated by a rather large and implausible star formation rate. 
This is partly due to inadequate accounting for specific massive star clusters nearby, such as in the Cygnus, Scorpius-Centaurus, and Orion regions. Interpreting remaining residuals, an even more local contribution to \Al emission may be missing in our model, which is difficult to find in analyses that rely on high-latitude references to define a background reference. 
We find that 23-40\% of the emission arises from the \emph{inner Galaxy} . This implies a rather large fraction of the total Galactic \Al emission arises from regions other than the inner Galaxy, including, e.g., the OB associations in the solar vicinity as listed in the cluster catalogues\cite{Melnik:2017}.

In any case, the spatial extent of \Al from and around massive-star clusters appears as our major systematic uncertainty. While superbubbles extending to sizes up to kpc are clearly required by observations, modelling these in representative morphology remains as a future task; here we adopted a simple homogeneous and spherical distribution isotropically extended around the cluster center with size increasing with age of the cluster. But the radial distribution within a superbubble obviously may be different, and morphologies of superbubbles deviate from sphericity if denser interstellar matter constrains expansion, or if galactic shear from galacticentric rotation and spiral-arm dynamics act during the expansion of older superbubbles.

\begin{figure} 
\centering
  \includegraphics[width=0.9\textwidth]{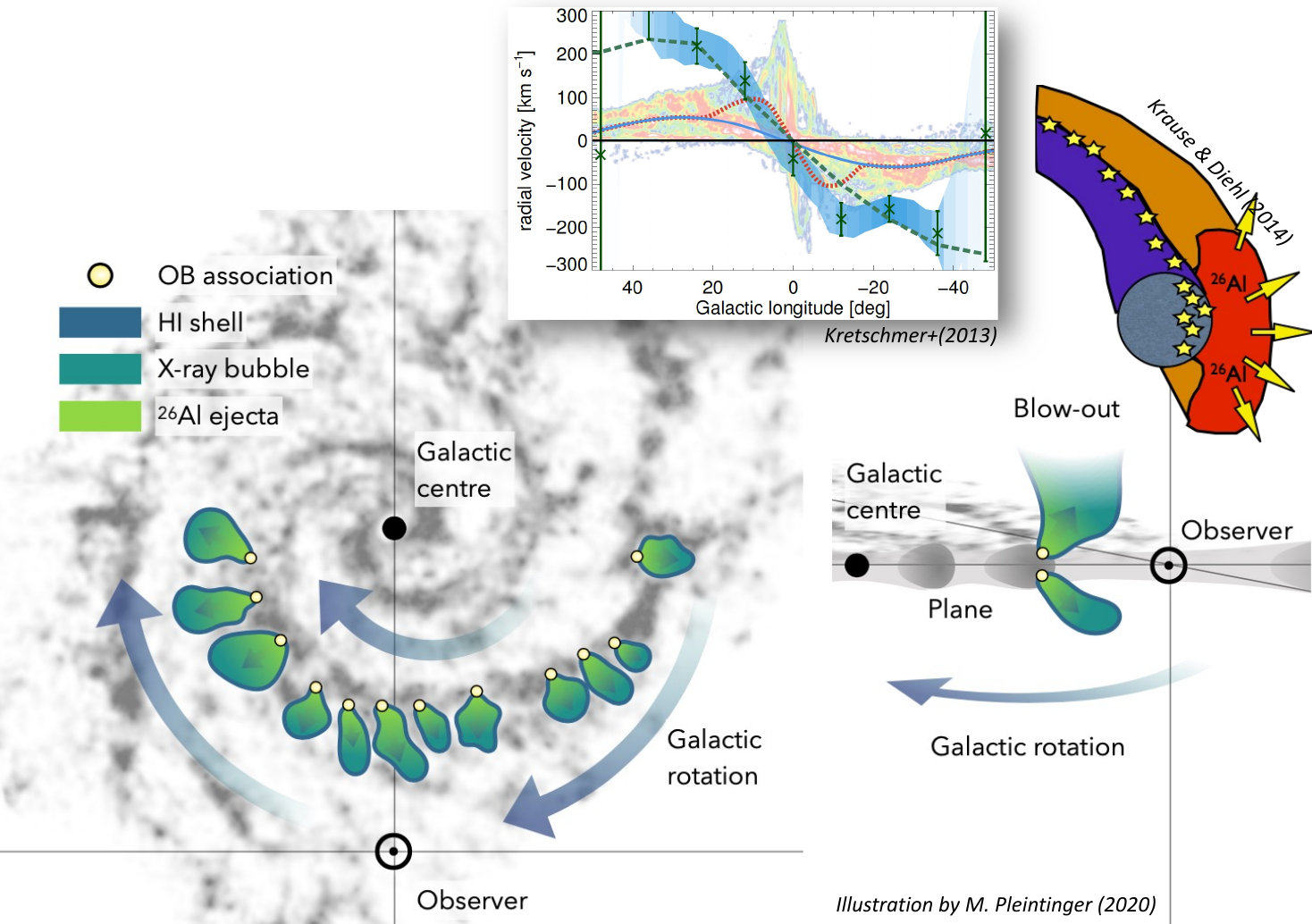}
  \caption{Illustrations of the spreading of nucleosynthesis ejecta from massive stars, as interpreted and understood from \Al emission measurements, and also supported from simulations and observations of other galaxies that can be viewed face-on (see text for details). }
  \label{fig:ejecta-spreading}
\end{figure} 

Beyond, as indicated by the trend of models to underpredict \Al as observed, it is quite possible that our model has other systematic deficits: The source yields may just be off from reality for the massive stars in the model. And/or there may be sources that have been overlooked. These could be a variety, from massive AGB stars\cite{Karakas:2017a}  and novae\cite{Jose:2020} to binary-system modifications of stellar evolution\cite{Zapartas:2017}. 

The best perspective of learning seems to be to measure in great detail and many observables the most-nearby massive star groups, in terms of their stellar population with well-determined ages and numbers, but also in terms of the interstellar surroundings in shapes and density and \Al emission. In such cases, the systematic uncertainties of our bottom-up model can be substantially reduced by consistency constraints from multiple messengers, and \Al data can contribute what nucleosynthesis and flow of ejecta in interstellar space can tell. 
The Orion region is such a nearby region, well observed and sufficiently far away from a potentially disturbing Galactic background, as discussed above. 
The Local Bubble with its recently consolidated origin from star clusters of the Scorpius-Centaurus region\cite{Alves:2018,Krause:2018,Zucker:2022} is another such nearby region, where now much is well-known about the stellar population and about the morphologies of dense interstellar matter. \Al emission may contribute important information about the hot and tenuous component of the interstellar medium, extending constraints from X-ray emission, and adding radioactivity as a special tracer to identify ejecta flows. 
Connecting these studies to the measurements\cite{Wallner:2021} of radioactive $^{60}$Fe and $^{244}$Pu and simulations\cite{Breitschwerdt:2016,Schulreich:2023} of ejecta transport in models add another tantalizing perspective for the study of ejecta transport from sources of nucleosynthesis towards next-generation star formation\cite{Diehl:2022}.

\section*{Acknowledgments}

R.D. acknowledges support from the Deutsche Forschungsgemeinschaft (DFG, German Research Foundation) under its Excellence Strategy, the Munich Clusters of Excellence \emph{Origin and Structure of the Universe} and \emph{Origins} (EXC-2094-390783311), and by the EU through COST action ChETEC CA160117.
 The {\it INTEGRAL}/SPI project has been completed under the responsibility and leadership of CNES;
  we are grateful to ASI, CEA, CNES, DLR, ESA, INTA, NASA and OSTC for support of this ESA space science mission.

\section*{References}
\bibliographystyle{unsrt}    

\end{document}